\documentclass[preprint2]{aastex}

\renewcommand\ion[2]{#1$\;${\scshape{#2}}}%

\shorttitle{Departure from the equilibrium in flaring plasma}
\shortauthors{Kawate et al.}

\begin{document}

\title{Departure of high temperature iron lines from the equilibrium state in flaring solar plasmas}

\author{T. Kawate\altaffilmark{1}, F. P. Keenan\altaffilmark{1}, and D. B. Jess\altaffilmark{1}}
\affil{Astrophysics Research Centre, School of Mathematics and Physics, Queen's University Belfast,
              Belfast, BT7 1NN, UK}
\email{t.kawate@qub.ac.uk}

\begin{abstract}
  The aim of this study is to clarify if the assumption of ionization equilibrium and a Maxwellian electron energy
  distribution is valid in flaring solar plasmas.
  We analyse the 2014 December 20 X1.8 flare, in which the \ion{Fe}{xxi} 187~\AA, \ion{Fe}{xxii} 253~\AA, \ion{Fe}{xxiii} 263~\AA\ and \ion{Fe}{xxiv} 255~\AA\ emission lines were simultaneously observed by the EUV Imaging Spectrometer onboard the Hinode satellite. Intensity ratios among these high temperature Fe lines are compared and departures from isothermal conditions
  and ionization equilibrium examined. 
Temperatures derived from intensity ratios involving these four lines show significant discrepancies at the flare footpoints in the impulsive phase,  and at the looptop in the gradual phase.  Among these, the temperature derived from the 
\ion{Fe}{xxii}/\ion{Fe}{xxiv} intensity ratio is the lowest, which cannot be explained if we assume a Maxwellian electron distribution and ionization equilibrium, even in the case of a multi-thermal structure.  
  This result suggests that the assumption of ionization equilibrium and/or a Maxwellian electron energy distribution can be violated in evaporating solar plasma around 10~MK.
\end{abstract}


\section{Introduction}

Solar flares are one of the most important phenomena to investigate the processes of energy development and its release in the solar atmosphere. Magnetic reconnection \citep{shib11} is now widely accepted as the energy release mechanism of solar flares both theoretically (e.g. \citealp{carm64}) and observationally (e.g. \citealp{tsun92}). However, the rate of energy transformation from magnetic into thermal, nonthermal and/or kinetic energy is still unknown. The derivation of 
physical parameters for the energetics is crucial for answering this question.

Spectroscopic observations are a powerful tool for diagnosing the physical parameters of the plasma.
Temperature and density diagnostics are, in most instances, based on the assumption of ionization equilibrium and a Maxwellian electron energy distribution. However, soft X-ray spectroscopic observations indicated that
the ion temperatures derived from satellite transitions or line widths are sometimes lower than the electron temperatures
during early flare stages \citep{dosc87b,kato98}. 
This indicates a thermal decoupling of these species, and the long collisional timescales have implications for other collisionally-dominated processes such as the ionization state and the electron distribution.
Emissivities under non-equilibrium ionization conditions due to heating and cooling processes during flares have also been investigated via numerical simulations \citep{brad03,real08}. The timescale to 
achieve ionization equilibrium depends on the electron density \citep{brad09,smit10}, and the non-equilibrium ionization state may not be negligible in both the energy release site \citep{imad11} and the evaporated plasma \citep{brad06}.

Non-Maxwellian distributions have been discussed primarily in the context of 
temperature diagnostics using soft X-ray satellite lines that are not affected by 
ionization processes \citep{gabr79, seel87}. Such non-Maxwellians have been employed as 
diagnostics of nonthermal electrons, and UV emission lines have also been examined to assess if they allow the detection of
nonthermal electrons \citep{pinf99,feld08,dzif10,dudi14}.

Here we examine the interrelationship of the intensities of high temperature lines that may be strongly affected by non-equilibrium ionization both spatially and temporally. We investigate if 
the assumption of ionization equilibrium and a Maxwellian electron distribution is valid 
in 10$^{7}$~K solar plasma during an X-class flare, using spectra from the EUV Imaging Spectrometer (EIS; \citealp{culh07}) onboard 
the Hinode satellite \citep{kosu07}.
Our paper is laid out as follows. 
In Section 2 we investigate the characteristics of high temperature Fe lines observed by EIS in terms of temperature and density under Maxwellian distribution and ionization equilibrium conditions, while in Section 3 we analyse an X-class flare and show results of  intensity interrelationships for the Fe lines. Finally in Section 4 we discuss possible departures from thermal
equilibrium and present our conclusions.

\section{Characteristics of flare lines in the EIS observation}

\subsection{ \ion{Fe}{XXI}, \ion{Fe}{XXII}, \ion{Fe}{XXIII} and \ion{Fe}{XXIV} }

We first briefly examine the characteristics of high temperature Fe lines we 
have selected. CHIANTI version 8.0.1 \citep{dere97,delz15} was used to calculate
the line intensities, and we adopted the coronal abundances of \cite{schm12} and ionization fractions of \cite{brya09}, with a
Maxwellian distribution in the ionization equilibrium.

\begin{figure*}
\begin{center}
\epsscale{2.0}
  \plotone{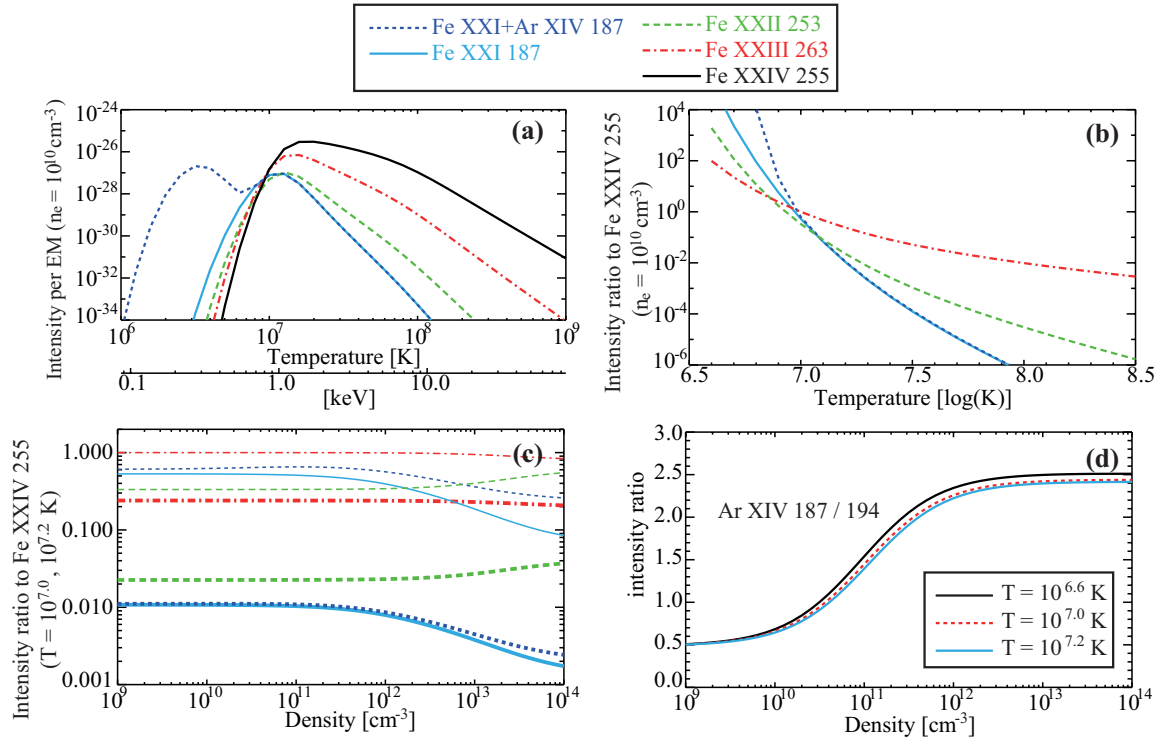}    
  \caption{Density and temperature dependence from CHIANTI of the solar emission lines considered in the present paper. (a) Contribution function of \ion{Fe}{xxi}+\ion{Ar}{xiv} 187~\AA \ (dark-blue dotted line), \ion{Fe}{xxi} 187~\AA \ (light-blue solid line), \ion{Fe}{xxii} 253~\AA \ (green dashed line), \ion{Fe}{xxiii} 263~\AA \ (red dot-dashed line), \ion{Fe}{xxiv} 255~\AA \ (black solid line). (b) Temperature dependence of intensity ratios relative to the \ion{Fe}{xxiv} 255~\AA\ line, with same line and colour coding
  as in (a). (c) Density dependence of intensity ratios relative to the \ion{Fe}{xxiv} 255~\AA\ line, with the same line and
  colour coding as in (a). Thick lines show intensity ratios at an electron temperature of log T$_{e}$ = 7.2, while thin lines show values for log T$_{e}$ = 7.0. (d) Density dependence of the  \ion{Ar}{xiv} 187~\AA/194~\AA \ intensity ratio at log T$_{e}$
  = 6.6 (black solid line), 7.0 (red dotted line) and 7.2 (blue solid line). } \label{fig:contfun}
\end{center}
\end{figure*}

Under the coronal approximation, most electrons are in the ground state and excitation is due to 
electron collisions. Thus, the line intensity depends on the electron collisional rate and the population of the upper level of the relevant transition along the line-of-sight. Hence if we derive the intensity ratio of 
two lines with significantly different excitation energies, this ratio will depend on both the 
electron energy distribution and the level populations. The electron energy distribution is, in most cases, assumed to be 
Maxwellian, i.e. a function of temperature. In addition, if we assume the plasma is in 
ionization equilibrium, the ratio of the line intensities is determined by temperature and column density. 
Figure~\ref{fig:contfun}(a) shows the contribution functions of the  lines considered here under the 
assumption of a Maxwellian distribution and ionization equilibrium. These lines are formed at similar temperatures around 10~MK.
We also show the temperature and density dependence of intensity ratios involving these lines in Figure~\ref{fig:contfun}(b) and (c), respectively. 
The temperature sensitivity is strong, while there is only a weak dependence on
 density. Therefore, if a hot plasma is assumed to be isothermal and in ionization equilibrium, the electron distribution is determined by the intensity ratios among pairs of these lines. However, if either assumption of an isothermal plasma 
 or ionization equilibrium is violated, the relationships in the figures no longer hold.
In a flaring region, \ion{Fe}{xxii} 253~\AA\ is unblended, and \ion{Fe}{xxiii} 263~\AA\ and \ion{Fe}{xxiv} 255~\AA\ only have some minor blended lines, while \ion{Fe}{xxi} 187~\AA\ is completely blended with \ion{Ar}{xiv}. 
There is another \ion{Ar}{xiv} line at 194~\AA \ in the EIS observation, although the ratio of these shows a density sensitivity around 10$^{10}$ to 10$^{12}$~cm$^{-3}$. 
Since there is uncertainty in the plasma densities, it is difficult to deblend the \ion{Fe}{xxi} + \ion{Ar}{xiv} 187~\AA\ feature, 
especially in a flaring region where the coronal density for a $10^7$~K plasma is around $10^{10}$ to $10^{12}$~cm$^{-3}$ \citep{dosc81,maso84,mill12}.

\subsection{Multi-thermal structures} \label{s:multitemp}

\begin{figure}
\begin{center}
\epsscale{.8}
  \plotone{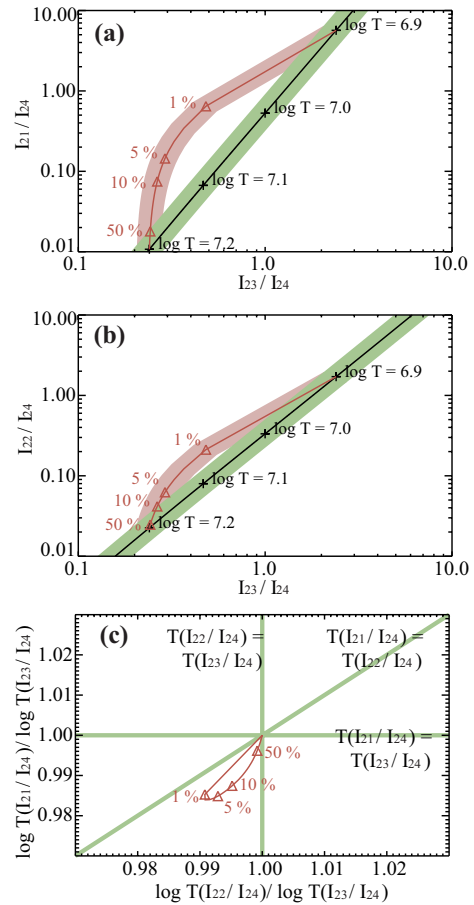}    
  \caption{Ratio-ratio plots calculated with CHIANTI of (a) $I_{21}/I_{24}$ versus $I_{23}/I_{24}$, 
  (b) $I_{22}/I_{24}$ versus $I_{23}/I_{24}$, (c) a ratio-ratio plot of $T(I_{21}/I_{24})/T(I_{23}/I_{24})$ versus $T(I_{22}/I_{24})/T(I_{23}/I_{24})$, where the ratios are defined in Section 2.2.  In (a) and (b), black lines show the isothermal state, 
  while green-filled regions show errors in intensity ratios under the isothermal state, estimated
  by assuming an uncertainty of $\pm$10\%\ in the adopted atomic data. 
  Red lines in (a), (b) and (c) indicate two-temperature models along the line-of-sight with temperatures of log T$_{e}$
  = 6.9 and 7.2. Triangles mark the fractions of the emission measure at log T$_{e}$ = 6.9 of 1, 5, 10 and 50\%. 
  The red-filled regions in (a) and (b) show errors in the intensity ratios under two-temperature conditions, once again estimated by adopting
  a $\pm$10\%\ uncertainty in the atomic data.} \label{fig:modelmt}
\end{center}
\end{figure}

A multi-thermal structure along the line-of-sight will result in a departure from the isothermal assumption in the optically thin solar corona. 
This has been discussed in earlier studies using 
differential emission measure analyses \citep{grah13,flet13}. Here we examine intensity ratios involving 
high temperature Fe lines in a multi-thermal structure in which each layer is assumed to have a Maxwellian distribution in
ionization equilibrium. Henceforth we denote the intensities of \ion{Fe}{xxi} 187~\AA , \ion{Fe}{xxii} 253~\AA , \ion{Fe}{xxiii}
263~\AA , and \ion{Fe}{xxiv} 255~\AA\ as $I_{21}$, $I_{22}$, $I_{23}$ and $I_{24}$, respectively. To simplify the 
analysis, we assume that the temperature structure consists of 
two components at log T$_{e}$ = 6.9 and 7.2 close to the peaks of the contribution functions of \ion{Fe}{xxi} to \ion{Fe}{xxiv}, and we vary the fraction of emission measures between the two temperatures. Figure~\ref{fig:modelmt}(a) and (b) show ratio-ratio plots involving the intensities of the four Fe lines, for 
different relative fractions of the emission measures. The total overall intensity arises from 
regions which have the larger fluxes in the lower ionized species, regardless of the relative 
fractions of the emission measures. To examine the relationships among the intensities simultaneously, 
 we also show ratio-ratio plots for temperatures from the line ratios in Figure 2(c).
 The figure suggests that, in the case of the two-thermal model log T$_{e}$ = 6.9 and 7.2, $T(I_{21}/I_{24}) < T(I_{22}/I_{24}) < T(I_{23}/I_{24})$ is always valid. This result comes from the curvature of the iso-thermal relationship among the line ratios shown in Figure~\ref{fig:contfun}(b). Hence, even if we examine the relationships at different temperatures, 
 $T(I_{21}/I_{24}) < T(I_{22}/I_{24}) < T(I_{23}/I_{24})$ is always valid under conditions of 
 ionization equilibrium and a Maxwellian distribution.

\section{Data analysis and results}

\subsection{Overview of observations}

Our observational dataset consists of an X1.8 class flare, which occurred in active region NOAA12242 on 2014 December 20. The GOES soft X-ray flux reached its maximum at 00:28 UT, and the location of the active region was S19W29 in the solar coordinate system. 
This flare was simultaneously observed by Hinode/EIS, the Atmospheric Imaging Assembly (AIA; \citealp{leme12}) onboard the Solar Dynamics Observatory (SDO; \citealp{pesn12}), and the Nobeyama Radio Polarimeter (NoRP; \citealp{tori79,naka85}) from the impulsive phase to the decay phase. NoRP observed microwave emission, which during solar flares mainly 
originates from semi-relativistic electrons in a flare loop via 
gyro-synchrotron emission. Hence, we can determine the time 
when the nonthermal electrons were created and the evaporated plasma filled the loop. 

EIS observations were 
performed in a slit scanning mode with a 2$^{\prime\prime}$ wide slit  and 3$^{\prime\prime}$
step size, and at a raster cadence of 534~s. The exposure time was 5~s, and the number of steps was 80 for one raster. Window height along the slit was 304 pixels with 
spatial sampling of 1$^{\prime\prime}$ pixel$^{-1}$. The field-of-view of the spatial range was therefore 304$^{\prime\prime}$ 
along the slit (north-south) and 240$^{\prime\prime}$ along the raster (west to east), centred at 
(445$^{\prime\prime}$, --263$^{\prime\prime}$).
EIS selected 15 spectral windows during these observations, and in our study we 
focused on the \ion{Fe}{xxi} 187 \AA , \ion{Fe}{xxii} 253 \AA , \ion{Fe}{xxiii} 263 \AA\ and \ion{Fe}{xxiv} 255 \AA \ lines, whose typical formation temperatures are about 10~MK.

\subsection{Calibration of spectral data}

We calibrated intensities of the EIS data by the following procedures. 
First, we ran \textit{eis\_prep} to subtract dark current, remove hot/warm pixels by cosmic rays, and calibrate the photometry using the laboratory data \citep{lang06}. Through this process we obtained level-1 data. 
Second, we ran \textit{eis\_wave\_corr\_hk} to correct the spatial offset in wavelength due to 
the orbital variation of the satellite \citep{kami10}. 
Third, we corrected the post-flight sensitivity of the absolute calibration by using the \textit{eis\_recalibrate\_intensity} function \citep{warr14}.
Fourth, we co-aligned spatial pixels along the wavelength direction by using 
 \textit{eis\_ccd\_offset} \citep{youn09}. 
The instrumental line FWHM for a slit width of 2$^{\prime\prime}$ in EIS is typically 62~m\AA\ \citep{brow08}, which the thermal FWHM is given by $2\sqrt{\ln 2 (kT/M_i)}$ in velocity unit, where $k$ is Boltzmann's constant, $T$ the temperature, and $M_i$ the mass of the ion. In the case of these Fe lines at their formation temperatures ($\sim 10^7$~K), this yields thermal FWHMs of 91~km s$^{-1}$, corresponding to 57~m\AA\ at 187~\AA\ and 80 m\AA\ at 263~\AA. Therefore, we cannot resolve lines within about $\pm 50$~m\AA \ of the high temperature Fe transitions. Also, during a 
flare these lines can be both red- and blue-shifted, with Doppler velocities of typically about 30 and 200~km s$^{-1}$, respectively \citep{mill09,hara11}, corresponding to 125--176~m\AA\ for these lines. 
Since \ion{Fe}{xxi} and \ion{Ar}{xiv} at 187~\AA \ are completely blended as discussed previously, we estimated
an upper limit for the \ion{Fe}{xxi} intensity by determining a lower limit for \ion{Ar}{xiv}, using the
measured \ion{Ar}{xiv} 194~\AA\ flux and the theoretical \ion{Ar}{xiv} 187~\AA/194~\AA\ ratio from CHIANTI.

To determine the continuum level accurately, we fitted lines in the same window simultaneously 
with a multi-Gaussian function using the MPFIT procedure \citep{mark09,more78}. Particularly in 
flare kernels, each line may have multiple components in one pixel~\citep{asai08}, so  we used a two-Gaussian function for each high-temperature Fe line to measure accurate intensities. Pixels in which intensities were less than 2$\times 10^3$~erg{\,}cm$^{-2}${\,}s$^{-1}$\AA $^{-1}${\,}sr$^{-1}$ were removed from the fitting. The number of Gaussian functions was six for the 188~\AA\ window,
four for 253~\AA, five for both 263~\AA\ and 255~\AA, and seven for 194~\AA . The \ion{Fe}{xii}, \ion{Fe}{xi} and \ion{O}{iv} ions in the 188, 188 and 253~\AA\ windows, respectively, each emit two lines in the same window. We 
assumed that each line pair has the same Doppler velocity and a fixed intensity ratio determined from CHIANTI. 
 There were several hot/warm pixels that were not flagged in the \textit{eis\_prep} procedure, and we removed these from the fitting manually.

\begin{figure*}
\begin{center}
\epsscale{2.0}
  \plotone{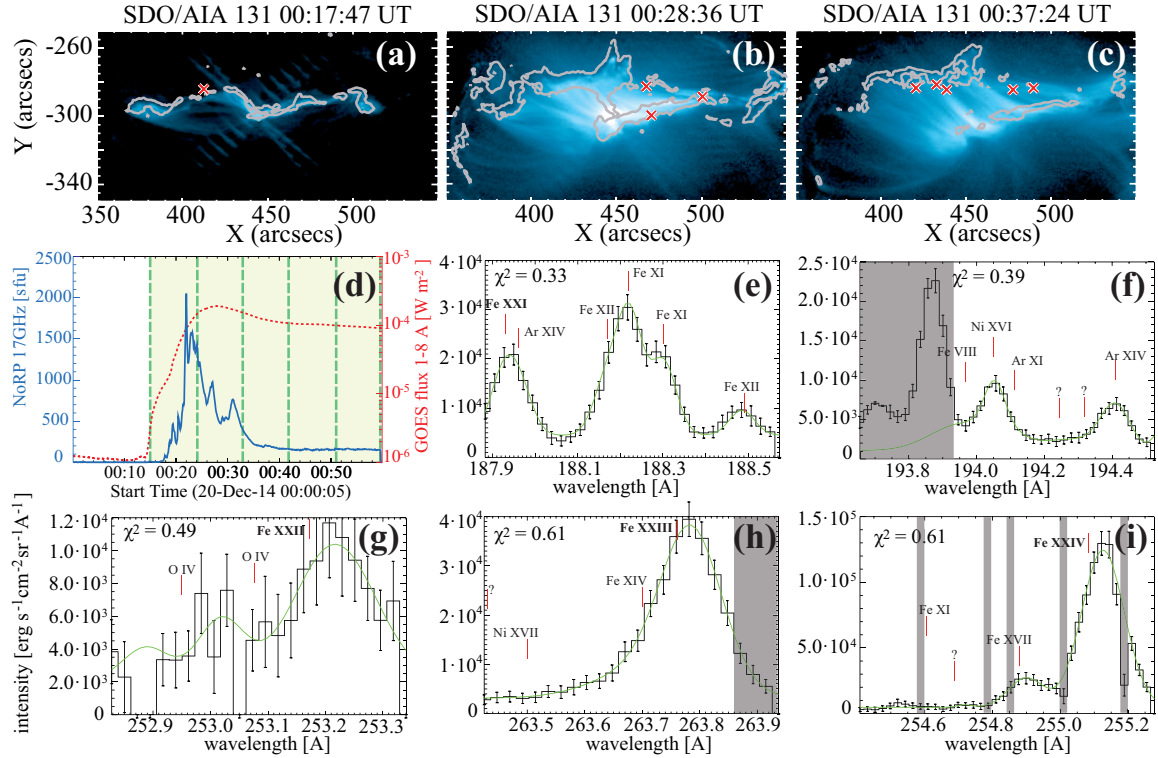}    
  \caption{AIA~131~\AA \ images of the 2014 December 20 solar flare during raster periods of (a) 00:15--00:24~UT, (b) 00:24--00:33~UT, and (c) 00:33--00:41, with the time when each image was observed at the top of the panel. The red crosses indicate 
  the position where the temperature shows a departure from isothermal. Grey contours show the AIA~1700~\AA \ image observed 6~s before each AIA~131~\AA \ one. (d) Lightcurves of the 17~GHz (blue thick line) and 1--8~\AA \ (red dotted line) emission observed by NoRP and GOES, respectively. Sets of raster exposures with EIS are shown as green dashed lines, and yellow-filled regions
  indicate when EIS flare data are available. (e)--(i) Example 
  spectra where the intensity relationship between $T(I_{22}/I_{24})$ and $T(I_{23}/I_{24})$ showed significant departures from  isothermal. Green solid curves show the best-fit function of the multi-Gaussian, and the $\chi ^2$ value for each is given in the 
  top-left portion of each plot. Emission lines included in the multi-Gaussian fitting are labelled,
  with high temperature Fe lines indicated with a bold font. Grey-filled regions show wavelengths we removed manually from the fitting due to warm pixels.} \label{fig:spectra}
\end{center}
\end{figure*}

 We determined the intensity of each line by integrating the Gaussian functions centered from --74 to +100~km{\,}s$^{-1}$ around each high temperature Fe line, corresponding to --46 to +62~m\AA\ at 187~\AA\ and --65 to +88~m\AA \ at 263~\AA . 
 This velocity limit is determined by the edge of the wavelength window of 187~\AA\ in the EIS data. Spatial pixels included in this analysis were limited by the following criteria: (i) the continuum intensity obtained by the fitting has a positive value in all five wavelength windows; (ii)
the reduced $\chi ^2$ of the fitting for \ion{Fe}{xxii}, \ion{Fe}{xxiii} and \ion{Fe}{xxiv} is less than 3; 
(iii) we only include the field-of-view spanning 350$^{\prime\prime}$ to 550$^{\prime\prime}$ in the east-west axis and 
--310$^{\prime\prime}$ to --260$^{\prime\prime}$ in the north-south axis, i.e. only regions around the flare. As a result, 633 sets of spectra were obtained. Examples of our fitting procedures are shown in Figure~\ref{fig:spectra}.

\subsection{Intensity ratios}

We calculated intensities of the Fe lines for a Maxwellian distribution and ionization equilibrium by changing temperature in CHIANTI. The grid points of temperature were from log T$_{e}$ = 6.6 to 7.6 in steps of 0.1 dex, with intermediate values
interpolated by a spline function. Ratios were calculated at a single density of N$_{e}$ =
10$^{10}$~cm$^{-3}$, as their dependence on density is small, changing by less than 6\%\ for densities
up to 10$^{11}$~cm$^{-3}$, smaller than the expected accuracy of the calculations given errors in the 
atomic data of 10\%\ \citep{chid05,delz05}. If the 
actual density is greater than 10$^{11}$~cm$^{-3}$, the  line that is most affected by high density is \ion{Fe}{xxi} 187~\AA,
and the derived temperature from this will be  overestimated. At N$_{e}$ = 10$^{11}$~cm$^{-3}$, the  
overestimation of the logarithmic temperature derived from $I_{21}/I_{24}$ increases with T$_{e}$,
but is only 
0.01 and 0.02 dex higher at log T$_{e}$ = 6.9 and 7.2, respectively. 

\begin{figure}
\begin{center}
\epsscale{.8}
  \plotone{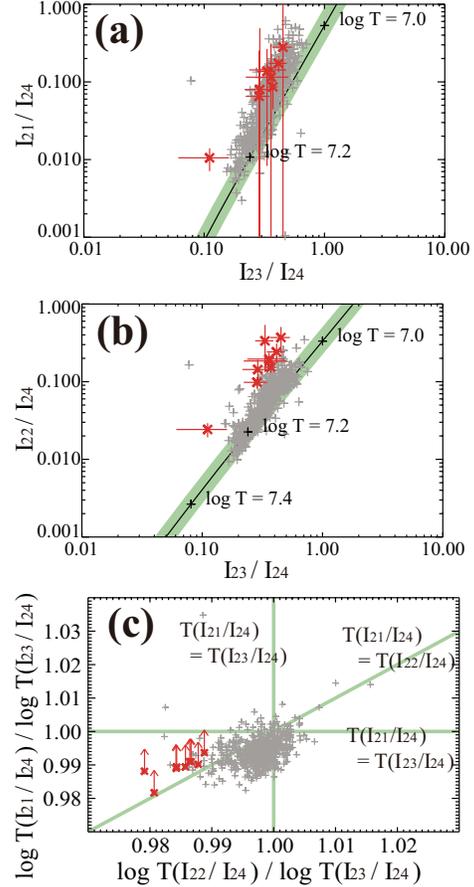}    
  \caption{(a) and (b) show ratio-ratio plots calculated with CHIANTI for 
  $I_{22}/I_{24}$ versus $I_{23}/I_{24}$ and $I_{21}/I_{24}$ versus $I_{23}/I_{24}$, respectively. Black lines show the ratio-ratio relationship under Maxwellian and ionization equilibrium conditions. Green-filled regions show the error in 
  the equilibrium intensity ratio assuming a $\pm$10\%\ error in the adopted atomic data. 
  Grey crosses show observed intensity ratios derived from pixels with well-fitted profiles ($\chi ^2 <3$). Red crosses
  with error bars are measured intensity ratios that are more than 1-$\sigma$ from the equilibrium state. (c) shows the temperature relationship among $T(I_{21}/I_{24})$, $T(I_{22}/I_{24})$ and $T(I_{23}/I_{24})$ derived from the observed intensity ratios. Red crosses  are data points that are more than 1-$\sigma$ from the equilibrium state between $T(I_{22}/I_{24})$ and $T(I_{23}/I_{24})$. Green lines show the relationships $T(I_{21}/I_{24})=T(I_{22}/I_{24})$, $T(I_{21}/I_{24})=T(I_{23}/I_{24})$, and $T(I_{22}/I_{24})=T(I_{23}/I_{24})$ under the isothermal assumption. } \label{fig:ratio} 
\end{center}
\end{figure}

As discussed in Section \ref{s:multitemp}, if the plasma is Maxwellian and in ionization equilibrium, the temperatures derived from line ratios
should show the relationship $T(I_{21}/I_{24}) < T(I_{22}/I_{24}) < T(I_{23}/I_{24})$.  However,
if the plasma does not obey these conditions, the derived temperatures may
not follow this relationship.
Figure~\ref{fig:ratio}(a) and (b) show the ratio-ratio relationships for the observations during 00:15--00:41~UT, with significant data points that lie more than 1 $\sigma$ from the equilibrium in the $I_{22}/I_{24}$ -- $I_{23}/I_{24}$ relation emphasised. As noted previously, the values of $I_{21}/I_{24}$ are upper limits. The corresponding observational 
points in the ratio-ratio diagram are displayed in Figure~\ref{fig:ratio} (c). We obtain two results from these plots. First, 
9 out of the 633 pixels in the flaring region show significant departure from the isothermal and ionization equilibrium 
conditions in the $I_{22}/I_{24}$ -- $I_{23}/I_{24}$ relation. 
Second, all pixels that show such a significant departure have a temperature relationship of $T(I_{22}/I_{24})<T(I_{23}/I_{24})$ and $T(I_{22}/I_{24})<T(I_{21}/I_{24})$. The mean value of $\log T(I_{22}/I_{24})/\log T(I_{23}/I_{24})$ is 0.985$\pm$0.001, while the lower limit of the mean value of $\log T(I_{21}/I_{24})/\log T(I_{23}/I_{24})$ is 0.989$\pm$0.001 among intensity ratios that show
significant departures in the rasters. From Figure~\ref{fig:ratio}, the highest temperature that shows a departure is log T$_{e}$
= 7.2, so that the above temperature relationship does not change even considering the case of an electron density of 
10$^{12}$~cm$^{-3}$.

 To further assess our results, we investigate in Figure~\ref{fig:spectra}(a), (b), and (c) the spatial position where the intensity ratio shows a significant departure from equilibrium.  
 In the figures, AIA 1700~\AA \ images are also plotted as a reference for the chromospheric flare footpoints.
 Comparing with the timing of the impulsive phase shown in Figure
 \ref{fig:spectra}(d), the significant departure appears at the footpoint in the impulsive phase, while in the gradual phase the departure appears mainly in the looptop. We plot in Figure\ref{fig:spectra}(e)--(i) one set of spectra from the pixel which shows a significant departure.
All of the spectra are well fitted using the multi-Gaussian function, producing maximum errors of $\chi ^2<0.61$.

\section{Discussion and summary}

We have examined the intensity relationships among Fe
lines observed in an X-class flare. For 9 out of 633 pixels, the temperatures derived from the intensities show departure from the isothermal and ionization equilibrium conditions. Temperature dependencies of $T(I_{22}/I_{24})<T(I_{23}/I_{24})$ and $T(I_{22}/I_{24})<T(I_{21}/I_{24})$ were found, suggesting that the assumption of a 
Maxwellian electron distribution and/or ionization equilibrium is violated. Pixels where the intensities showed a significant departure from the isothermal condition in the $I_{22}/I_{24}$ - $I_{23}/I_{24}$ relation are located at the footpoint in the impulsive phase, and looptop in the gradual phase.

The number of pixels that show departures from isothermal and ionization equilibrium conditions is as small as 1.4\% compared
to that of valid pixels in the entire flaring region. Therefore, we could conclude that ionization equilibrium is valid in most cases 
within the timescale of the EIS exposures. However, all pixels that show a departure from thermal equilibrium have the same 
temperature relationship, which implies the same physical processes are occurring in the region. 
The number of pixels is highly influenced by the timing of the slit exposure, errors in the observations, 
and the validity of the assumption in the models.
Nevertheless, we can also conclude that the assumption of isothermal and ionization equilibrium conditions is not valid in some cases.
Significant departures from this assumption can be explained by the following. 
The departure from equilibrium conditions appeared at the footpoint in the impulsive phase and looptop in the late gradual phase, 
suggesting that the departure arises  along the path of evaporation. 
At the footpoints of the impulsive phase, the non-thermal tail under a non-Maxwellian electron distribution would favor the creation of more strongly ionized species, as well as rapid heating due to the evaporated plasmas. The apparent temperatures among these line ratios are always overestimated under non-equilibrium ionization and a non-Maxwellian distribution. Even examining pure non-equilibrium ionization, it takes about 10$^3 (10^9/N_e)^{-1}$ s to reach ionization equilibrium for \ion{Fe}{xxiv} \citep{brad09,imad11}. Since \ion{Fe}{xxi} starts to ionize earliest among these species, 
$T(I_{21}/I_{24})$ is higher than  $T(I_{22}/I_{24})$ and  $T(I_{23}/I_{24})$ in the heating phase.
This may explain the observed temperature relationships if $T(I_{21}/I_{24}) > T(I_{23}/I_{24})$ is valid, although we cannot confirm this since we only provide a lower limit to $T(I_{21}/I_{24})$. If $T(I_{21}/I_{24}) > T(I_{23}/I_{24})$ is not valid, a non-Maxwellian electron distribution may couple with a non-equilibrium ionization in a multi-thermal structure, and we would need detailed numerical simulations to understand this fully.
On the other hand, at the looptop in the gradual phase, the evaporated plasmas fill the flare loop and radiative cooling dominates in the temperature evolution.  More highly ionized species are over populated, and the temperature relationship is  $T(I_{21}/I_{24})<T(I_{22}/I_{24})<T(I_{23}/I_{24})$, which cannot be distinguished from the relationship under ionization equilibrium, and multi-thermal structures cannot explain the observed result. 
An explanation for the observed result would be coupling of the high energy tail in the electron distribution, i.e., a non-Maxwellian distribution with non-equilibrium ionization. 
However, it is difficult to solve the inverse problem, (i.e., determine the degree of non-equilibrium ionisation or extent of non-thermal structures) solely from high-temperature line ratios.
Further studies of combined models for the simultaneous
evolution of electron distribution and non-equilibrium ionization, and employing better sensitivity with higher cadence observations, are needed to explain this phenomenon.

\acknowledgments
The authors are grateful to Dr. R. Milligan, Dr. S. Imada, and Dr. H.~E. Mason for fruitful discussions, and also appreciate the anonymous referees for the comments to improve the paper.
T.K. wishes to thank the UK Science and Technology Facilities Council (STFC) for funding. 
D.B.J. is grateful to STFC for an Ernest Rutherford Fellowship, in addition to a dedicated research grant that allowed this work to be undertaken.
Hinode is a Japanese mission developed and launched by ISAS/JAXA, with NAOJ as domestic partner and NASA and STFC (UK) as international partners. It is operated by these agencies in co-operation with ESA and NSC (Norway). 
CHIANTI is a collaborative project involving George Mason University, the University of Michigan (USA) and the University of Cambridge (UK). 
{\it Facilities:} \facility{Hinode(EIS)}, \facility{NoRP}, \facility{CHIANTI}.

\bibliographystyle{apj}

\bibliography{ref}

\end{document}